\documentclass{ws-procs975x65}

\begin{document}

\def\lsim{\lower.5ex\hbox{$\; \buildrel < \over \sim \;$}}
\def\gsim{\lower.5ex\hbox{$\; \buildrel > \over \sim \;$}}

\title{Neutrino asymmetry in presence of gravitational interaction}

\author{Banibrata Mukhopadhyay
\footnote{\uppercase{W}ork partially
supported by grant 80750 of the \uppercase{A}cademy of \uppercase{F}inland.}}
\address{Astronomy Division, P.O.Box 3000, University of Oulu,
FIN-90014, Finland; bm@cc.oulu.fi}

\maketitle

\abstracts{
Propagation of fermions in curved space-time generates a
gravitational interaction due to the coupling between spin of the
fermion and space-time curvature. This gravitational interaction,
which is an axial-vector appears as the CPT violating term in the
Lagrangian, 
can generate the neutrino asymmetry in Universe.
If the back-ground metric is spherically asymmetric, say, of rotating
black hole, the axisymmetric space-time of an expanding Universe, 
this interaction as well as the neutrino asymmetry is non-zero. 
}

A spinning test particle, fermion, gives rise to a gravitational interaction inherently
in the curved space-time through its propagation. Actually, in presence of the strong 
gravitational field, fermionic spin couples with the connection coefficients of the background 
space-time and produces this interaction. As the gravity effect reduces, the interaction
strength decreases too. Therefore in flat space-time, e.g. our Earth, this interaction practically
vanishes. Thus in Earth, fermions appear
as free, which is not the case in presence of significant background curvature.
When a fermion is considered to be propagating in the early phase of Universe and/or in the space-time
around {\it rotating} black holes, when the space-time curvature effect can not be neglected, 
this spin(fermionic)-spin(curvature) interaction may become significant. 

The generation of neutrino asymmetry in early Universe of an well known fact. However, apart from such  
an asymmetry in early Universe, one can ask about the possibility of an additional asymmetry in the
present era. Here I will show that this asymmetry may even arise in the present era when
neutrinos are propagating under a strong gravitational field, through the appearance of 
gravitational interaction mentioned above. This interaction may be responsible for further
neutrino asymmetry, in the early as well as present era when the space-time is non-flat.

The general metric for curved background can be given as
\begin{eqnarray}
ds^2=g_{00}dt^2-g_{ii}(dx^i)^2-g_{0i}dtdx^i-g_{ij}dx^idx^j,
\label{met}
\end{eqnarray}
where $i\rightarrow 1-3$.
Separating out different components of the metric I get 
\begin{eqnarray}
g_{\mu \nu}=
 \underbrace{\left[ \begin{array}{cccc} c_{  1} & 0 & 0 & 0 \\
                                             0 & c_{ 2} & 0 & 0 \\
                                             0 & 0 & c_{ 3} & 0 \\
                                             0 & 0 & 0 & c_{ 4}
                        \end{array} \right]}_{I} +  \underbrace{\left[ \begin{array}{cccc} 0 & c_{
5} & c_{ 6} & c_{ 7} \\
                                              c_{ 5} & 0 & 0 & 0 \\
                                              c_{ 6} & 0 & 0 & 0\\
                                             c_{7} & 0 & 0 & 0
                        \end{array} \right]}_{II} +  
\underbrace{\left[ \begin{array}{cccc} 0 & 0 & 0 & 0 \\
                                             0 & 0 & c_{ 8} & c_{ 9} \\
                                             0 & c_{ 8} & 0 & c_{ 10} \\
                                             0 & c_{ 9} & c_{ 10} & 0
                        \end{array} \right]}_{III}.
\label{metsep}
\end{eqnarray}
Therefore along with (\ref{met}) the most general Dirac Lagrangian density is\cite{mmp02}
\begin{eqnarray}
\hskip-1cm
{ L}=det(e)\left(i\overline{\psi}\gamma^a D_a\psi-m\overline{\psi}\psi\right),\hskip0.1cm
D_a=\partial_a-\frac{i}{4}\omega_{bca}\,\sigma^{bc},\hskip0.1cm
\omega_{bca}=e_{b\lambda}\left(\partial_a\, e^\lambda_c+
\Gamma^{\lambda}_{\gamma\mu}\,e^\gamma_c\, e^\mu_a\right),
\label{lag}
\end{eqnarray}
where $a\rightarrow 0-4$.
After expanding out and combining with its conjugate part, $L$ can be expressed as
\begin{eqnarray}
{L}=det(e)\overline{\psi}\left[\left(i\gamma^a \partial_a-m\right)+
\gamma^a\gamma^5B_a\right]\psi.
\label{lex}
\end{eqnarray}
The gravitational interaction part of $L$  can be identified as
\begin{eqnarray}
{L}_I=det(e)\overline{\psi}\gamma^a\gamma^5B_a\psi,\hskip0.3cm 
B_a=\epsilon^{dbc}\,_a\, e_{b\lambda}\left(\partial_d\, e^\lambda_c+\Gamma^{\lambda}_{\alpha\mu}\,
e^\alpha_c\, e^\mu_d\right).
\label{lint}
\end{eqnarray}
The interesting point is, this $L_I$, which is an axial vector multiplied by a gravitational 
four-vector potential, is odd under CPT transformation, if $B_a$ is unaffected. 
Therefore $L_I$ will have different sign for a left-handed particle and right-handed 
anti-particle. It should be noted that similar interactions are considered in the CPT violating
and string theories [e.g. works done 
by Kosteleck\'y and his collaborators\cite{kos}]. But here the terms come in the picture automatically.
Now if I expand the associated axial vector in $L_I$ in case of a neutrino
($\psi$) and corresponding anti-neutrino ($\psi^c$) (keeping in mind, according to the standard model the
neutrino is purely left-handed and the associated anti-neutrino is purely right-handed),
get\cite{sm}
\begin{eqnarray}
\overline{\psi}\gamma^a\gamma^5\psi=-\overline{\psi}_L\gamma^a\psi_L,\hskip0.4cm
\overline{\psi^c}\gamma^a\gamma^5\psi^c=\overline{\psi^c}_R\gamma^a\psi^c_R.
\label{axil}
\end{eqnarray}
Therefore the dispersion relations for neutrinos ($\nu$) and anti-neutrinos ($\overline{\nu}$) are
\begin{eqnarray}
E_{\nu,{\overline{\nu}}}=\sqrt{|\vec{p}|^2\pm 2(B_0p^0+B_ip^i)+B_aB^a-m^2},
\label{dis}
\end{eqnarray}
where `$+$' and `$-$' are for $\nu$ and $\overline{\nu}$ respectively. Clearly $E_{\nu}$ and
$E_{\overline{\nu}}$ are different and thus the difference in number density between neutrinos and
anti-neutrinos can be evaluated as
\begin{eqnarray}
\Delta n=\frac{g}{(2\pi)^3}\int d^3|\vec{p}|\left[\frac{1}{1+exp(E_\nu/T)}
-\frac{1}{1+exp(E_{\overline{\nu}}/T)}
\right].
\label{asy}
\end{eqnarray}
$\Delta n \neq 0$, only if $B_0\neq 0$. According to (\ref{metsep}) this is only
possible if $III\neq 0$. Therefore, to have a non-zero neutrino asymmetry under gravity
background, the space-time metric must have an off diagonal pure spatial component. 
It is now also understood, the presence of $B_0$ (gravitational scalar potential) is responsible 
to neutrino asymmetry to occur according to this formalism. 

The space-time in early Universe which is homogeneous and anisotropic, e.g. Bianchi Model II,
VIII and IX (e.g. see reference\cite{bsc90}), which are axially symmetric, can generate $B_0$ to non-zero and then neutrino
asymmetry. Also the space-time around a rotating black hole, Kerr geometry, which is also axially 
symmetric can give rise to non-zero $B_0$. The form of Kerr metric\cite{d00} expressed like equation 
(\ref{met}) is very suitable to understand this feature and compute $B_0$.

Let us consider a special case of space-time where $\vec{B}.\vec{p} <<B_0 p^0$, $B_a B^a <<1$
and $B_0 <<T$. Thus from (\ref{asy}), in an ultra-relativistic regime
$\Delta n\sim \overline{B_0} T^2$, where $\overline{B_0}$ indicates the integrated value of
$B_0$ over the space.
                                                                                              
If I consider the Bianchi II model with same scale factors in all directions,                                                
at $t\sim 10^{-12}$ sec when the electroweak era finishes or about to finish, $B_0\sim R(t)\lsim 10^{-6}$.
Then for $T\lsim 10^{-2}$ erg, $\Delta n \lsim 10^{-10}$.
In case of an accretion disk around black holes, $T\sim 10^{11}$ K $\sim 10$ MeV $ \sim 10^{-5}$ erg.
If the black hole parameters are such, those satisfy the approximation of the metric and
$\overline{B_0}\lsim 10^{-6}$ erg; $\Delta n \lsim 10^{-16}$.
In case of Hawking radiation bath, $T\sim 10^{-7}\left(\frac{M_\odot}{M}\right)$ K. The primordial black
holes of mass, $M>10^{15}$ gm, still exist today. The temperature of those black holes is,
$T>10^{11}$ K $\sim 10^{-5}$ erg. Therefore, if $\overline{B_0}=10^{-6}$, $\Delta n \gsim 10^{-16}$, depending on
the mass of the black holes. 
                                                                                                                                              
If there are $N_i$ number of $i$-kind black holes with corresponding curvature scalar
coupling as $\overline{B_0}_i$ and temperature $T_i$, the neutrino asymmetry for that particular
kind of black hole can be written as
\begin{equation}
\Delta n_i=10^{-10}\left(\frac{N_i}{10^6}\right)\left(\frac{\overline{B_0}_i}{10^{-6} erg}\right)
\left(\frac{T_i}{10^{-5} erg}\right)^2.
\label{neuasy}
\end{equation}
Thus the overall asymmetry for all kinds of black hole becomes
$\Delta n=\sum_{i} \Delta n_i$.
In Earth, $\overline{B_0}_{earth}\sim 10^{-40}$ erg, $T_{earth}\sim 10^{-14}$ erg. Then
$\Delta n_{earth}\sim 10^{-68}$, which is very small over the relic asymmetry and thus
one does not see its effect. However, in a laboratory of earth, if one is able to thermalise
the neutrino, asymmetry could be raised.

Therefore I propose, to generate the neutrino asymmetry under strong gravity, all the following
criteria have to be satisfied simultaneously as: (i) The space-time should be axially symmetric,
(ii) the interaction Dirac Lagrangian must have a CPT violating term which may be an
axial-four vector (or pseudo-four vector) multiplied by a curvature coupling four vector potential,
(iii) the temperature scale of the system should be large with respect
to the energy scale of the space-time curvature.

\end{document}